# Experimental Realization of Weyl Exceptional Rings in a Synthetic Three-Dimensional Non-Hermitian Phononic Crystal

Zheng-wei Li[1], Jing-jing Liu[1], Ze-Guo Chen[2], Weiyuan Tang[2], An Chen[1], Bin Liang[1]*, Guancong Ma[2]*, Jian-chun Cheng[1]

[1]*Collaborative Innovation Center of Advanced Microstructures and Key Laboratory of Modern Acoustics, MOE, Institute of Acoustics, Department of Physics, Nanjing University, Nanjing 210093, People's Republic of China*

[2]*Department of Physics, Hong Kong Baptist University, Kowloon Tong, Hong Kong, China.*

*Emails: liangbin@nju.edu.cn (B. L.); phgcma@hkbu.edu.hk (G. M.)

**Abstract:** Weyl points (WPs) are isolated degeneracies carrying quantized topological charges, and are therefore robust against Hermitian perturbations. WPs are predicted to spread to the Weyl exceptional rings (WERs) in the presence of non-Hermiticity. Here, we use a one-dimensional (1D) Aubry-Andre-Harper (AAH) model to construct a Weyl semimetal in a 3D parameter space comprised of one reciprocal dimension and two synthetic dimensions. The inclusion of non-Hermiticity in the form of gain and loss produces a WER. The topology of the WER is characterized by both its topological charge and non-Hermitian winding numbers. The WER is experimentally observed in a 1D phononic crystal with the non-Hermiticity introduced by active acoustic components. In addition, Fermi arcs are observed to survive the presence of non-Hermitian effect. We envision our findings to pave the way for studying the high-dimensional non-Hermitian topological physics in acoustics.

*Introduction.* Topological phenomena were discovered in condensed matter physics, and soon thereafter extended to photonic and phononic crystals (PCs)[1-4]. One important class of three-dimensional (3D) topological systems is the Weyl semimetal, with the existence of Weyl points (WPs) – isolated two-fold degeneracies at the linear crossings of two bands – as a defining characteristic[5]. A WP carries a quantized topological charge and is a source or sink of the Berry curvature flux. Thus

WPs must emerge in pairs and WPs of opposite charges are connected by a spectral excitation called a Fermi arc. WPs have been observed in photonics[6-10], acoustics[11-15], and various condensed matter systems[16-20]. The physics in the vicinity of a WP can typically be captured by a two-level Weyl Hamiltonian, which contains all the components of Pauli matrices. Such a mathematical form means that WPs are robust against Hermitian perturbations. However, perturbations that break Hermiticity can lead to entirely different scenarios. Non-Hermitian systems can give rise to a special kind of spectral "degeneracy" called an exceptional point (EP)[21-25], at which one or more state vector(s) become defective. The presence of non-Hermiticity can transform Hermitian degenerate points, such as a Dirac-like point or WP, into an exceptional ring, which is a continuous closed trajectory of EPs in the reciprocal space[26-29]. In particular, the realization of a Weyl exceptional ring (WER) in fully reciprocal space demands fine control of non-Hermitian parameters in a 3D crystal, which is experimentally challenging and is so far only realizable in a helical photonic waveguide array [29]. On the other hand, the recent development in the synthetic dimensions indicates that system parameters can be harnessed as degrees of freedom that map to new system dimensions, which opens a convenient route to study higher-dimensional physics using a system with fewer real dimensions[7,9,11,30-33].

This work presents an experimental realization of the WER in a hybrid synthetic-reciprocal space[34] implemented by a one-dimensional (1D) PC designed as a chain of coupled acoustic cavities with actively controlled loss and gain[35,36]. Our system is based on a 1D Aubry-Andre-Harper (AAH) model[37,38], in which both the hopping and onsite energy terms are modulated by two separate cosine functions. These modulations enforce two additional synthetic dimensions. We further introduce non-Hermiticity of the Hamiltonian by the inclusion of gain and loss, and a WER spawns from the WP. Meanwhile, Fermi arcs are found connecting the two WPs or two WERs with different Hermitian topological charges in the synthetic 3D parameter space. Our findings are verified in acoustic experiments. This work paves the way to explore high-dimensional non-Hermitian topological physics with low-dimensional PCs.

***Theory and design of system***. We start with the following AAH model Hamiltonian,

$$H(k_x, \xi_y, \xi_z) = \begin{bmatrix} \omega_-(\xi_z) & \kappa(k_x, \xi_y) \\ \kappa^*(k_x, \xi_y) & \omega_+(\xi_z) \end{bmatrix} + \begin{pmatrix} -i\gamma_1 & 0 \\ 0 & i\gamma_2 \end{pmatrix} - i\gamma_0 \sigma_0, \tag{1}$$

where

$$\omega_\pm(\xi_z) = \omega_0 \pm b_1 \cos \xi_z, \tag{2}$$

$$\kappa(k_x, \xi_y) = \kappa_+(\xi_y) + \kappa_-(\xi_y) e^{-ik_x a}, \tag{3}$$

with $b_1 = 50$, $b_2 = 0.5$, $a$ is lattice constant, $\omega_0$ is the pre-modulation onsite energy, $\kappa_+(\xi_y) = -\kappa_0[1 + b_2 \cos(\xi_y + \pi)]$ and $\kappa_-(\xi_y) = -\kappa_0(1 + b_2 \cos \xi_y)$ are the intra-cell and inter-cell coupling parameters, $k_x$ is the Bloch wavenumber. The two imaginary parameters, $-i\gamma_1$ and $i\gamma_2$, respectively represent additional loss and gain, which introduce non-Hermiticity. $-i\gamma_0$ denotes intrinsic onsite loss ($\gamma_0, \gamma_1, \gamma_2 \geq 0$) and $\sigma_0$ is an identity. $\xi_y$ and $\xi_z$ are two parameters that modulate the hopping and onsite real energy terms. Hence, the eigenstates of Eq. (1) are bundles on a 3D base manifold, in which $k_x$ identifies with $k_x + 2\pi/a$, $\xi_y$ with $\xi_y + 2\pi$, and $\xi_z$ with $\xi_z + 2\pi$. Thus the base manifold is isomorphic to a 3D torus, which is also isomorphic to a 3D Brillouin zone. Consequently, we can regard $\xi_y, \xi_z$ as two synthetic dimensions, which, together with the reciprocal dimension $k_x$, make the system described by Eq. (1) effectively a 3D periodic system.

In the Hermitian limit, one has $\gamma_1, \gamma_2 = 0$, and Eq. (1) gives two-fold degenerate points in the synthetic-reciprocal space at $(k_x, \xi_y, \xi_z) = \left(\pm\frac{\pi}{a}, \pm\frac{\pi}{2}, \pm\frac{\pi}{2}\right)$, as shown in Fig. 1a. We analyze the degenerate point at $\left(\frac{\pi}{a}, \frac{\pi}{2}, \frac{\pi}{2}\right)$. In its vicinity, the Hamiltonian is, retaining the linear terms and dropping $\gamma_0$,

$$H = \omega_0 \sigma_0 + d_x \sigma_x + d_y \sigma_y + d_z \sigma_z, \tag{4}$$

where $\sigma_x, \sigma_y, \sigma_z$ are Pauli matrices and $d_x = -2b_2\kappa_0(\xi_y - 0.5\pi), d_y = -\kappa_0(k_x a - \pi), d_z =$

$b_1(\xi_z - 0.5\pi)$. Equation (4) takes the form of a Weyl Hamiltonian[16], and the dispersion is linear in all directions.

We characterize the topology of the degenerate point by calculating its Berry charge. To do so, we integrate the Berry curvature over a surface enclosing the point (gray spherical shell in Fig. 2a)[27,28,39].

$$C_m = \frac{1}{2\pi} \int_{\partial\Omega} \vec{A}(\vec{\mu}) \cdot d\vec{S} \tag{5}$$

where $A_{mn}(\vec{\mu}) = i\langle \nabla_{\vec{\mu}}\psi_m(\vec{\mu}) | \times | \nabla_{\vec{\mu}}\psi_n(\vec{\mu}) \rangle$ is the local Berry curvature, with $\psi_{m,n}$ being the Bloch wavefunctions of the two bands and $\mu = k_x, \xi_y, \xi_z$. Our calculations show that $C = \pm 1$, confirming that the two-fold degenerate point is a WP[40]. The charges of the two WPs at $\left(\frac{\pi}{a}, \pm\frac{\pi}{2}, \frac{\pi}{2}\right)$ are opposite, indicating that they form a pair of source and sink for the Berry flux and must be connected by a Fermi arc, as will be shown later.

Next, we introduce non-Hermiticity by adding loss and gain $(\gamma_0, \gamma_1, \gamma_2)$ to the onsite terms of the Hamiltonian. The eigenfrequencies of Eq. (1) take the form

$$\widetilde{\omega}_{1,2} = \omega_0 - i\frac{2\gamma_0 + \gamma_1 - \gamma_2}{2} \pm \frac{1}{2}\sqrt{4|\kappa(k_x, \xi_y)|^2 - (\gamma_1 + \gamma_2 - 2ib_1\cos\xi_z)^2}. \tag{6}$$

In doing so, EPs are spawned from the WP. The WP morphs into a continuous closed trajectory on the $k_x\xi_y$ plane at $\xi_z = 0.5\pi$, at which the real and imaginary parts of the eigenvalues are identical, as shown in Fig. 1c and Figs. 1e, f. This trajectory is a WER. The shape of the WER is given by

$$(b_2 \cos \xi_y)^2 + \cos k_x [1 - (b_2 \cos \xi_y)^2] = \left(\frac{\gamma_1 + \gamma_2}{2\sqrt{2}\kappa_0}\right)^2 - 1. \tag{7}$$

Figures 1h, i show the bands on the $k_x\xi_z$ cut plane at $\xi_y = 0.5\pi$, which intersect the WER at the two points marked by the red stars.

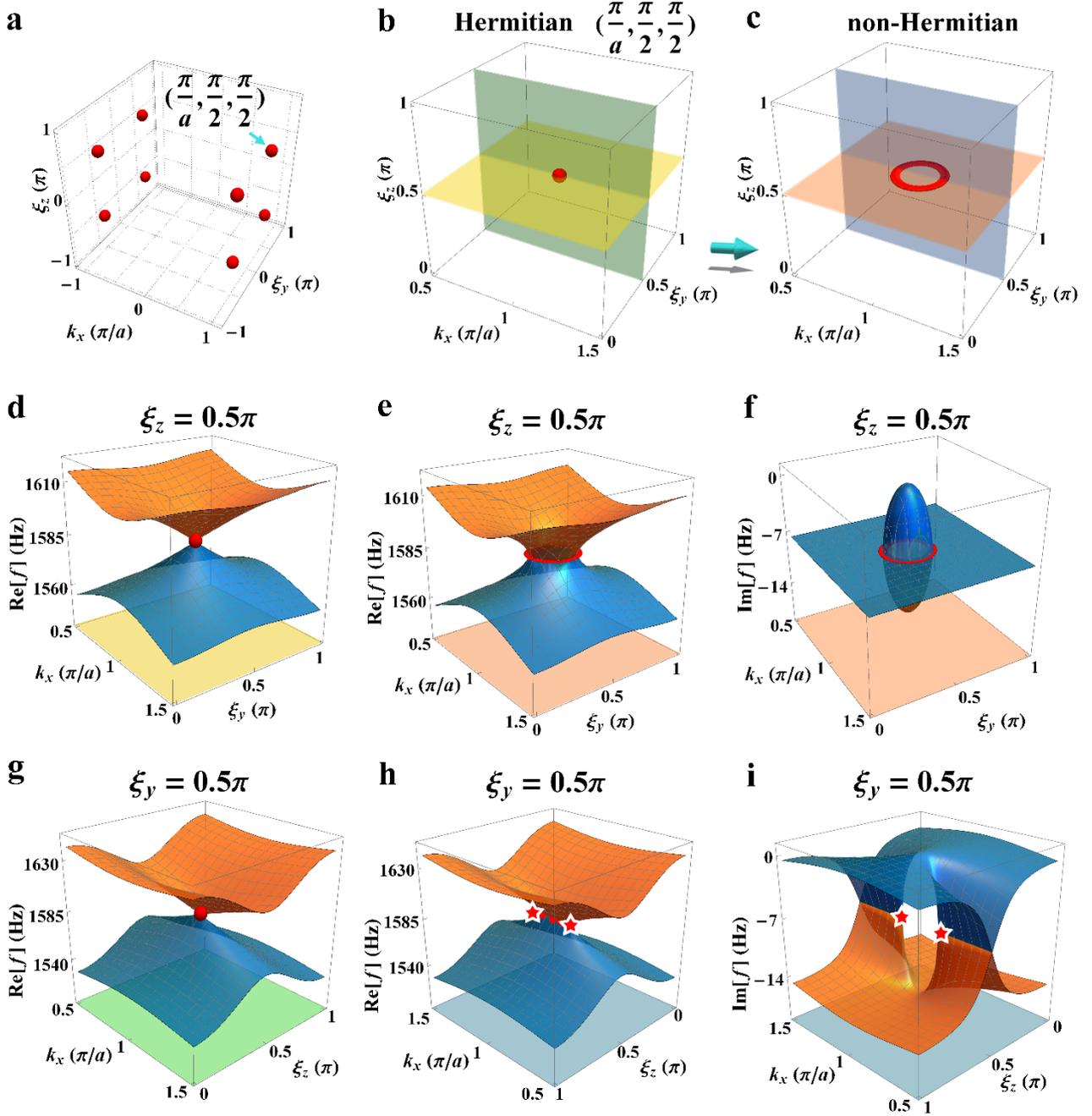

**Fig. 1 WPs and the WER in the synthetic parameter space and corresponding band structures.**
**a** WPs in the synthetic-reciprocal space. **b**, **c** Schematics of the WP located at $\left(\frac{\pi}{a}, \frac{\pi}{2}, \frac{\pi}{2}\right)$ evolving to the WER. **d** to **i** show the band structures on different cut planes as indicated by the colors. **d**, **g** Band structures in the Hermitian limit in the $k_x\xi_y$ plane at $\xi_z = 0.5\pi$ and $k_x\xi_z$ plane at $\xi_y = 0.5\pi$, respectively. The WP is marked by the red dot. **e**, **f** Real part and imaginary part of the spectrum in the $k_x\xi_y$ plane at $\xi_z = 0.5\pi$ in the non-Hermitian system. A WER is identified and marked by the red circles. **h**, **i** The same system viewed on $k_x\xi_z$ plane with for $\xi_y = 0.5\pi$.

The emergence of the WER gives rise to richer topological properties, which can be characterized at two different levels. First, similar to the WP, we can compute the Berry charge of the WER by integrating the Berry curvature over an enclosing surface shown in Fig. 2b. Our results show that $C = \pm 1$, implying that the WER retains the WP's topological charge, despite the inclusion of non-Hermiticity. This is further verified by the Berry curvature of two WPs and WERs at $\left(\frac{\pi}{a}, \pm\frac{\pi}{2}, \frac{\pi}{2}\right)$, as shown in Figs. 2c, d. The results also suggest that Fermi arcs survive the inclusion of non-Hermiticity as detailed in the following experimental section. These results agree with related theoretical calculations[27]. Second, the EPs comprising the WER also possess non-Hermitian topology. Unlike Hermitian degeneracies, the topological properties of an EP can be characterized by two methods: eigenvalues and eigenvectors[41,42]. The eigenvalue holonomy around an EP generates a winding number[42-44], given by $Y = \sum_{j,j'=1}^{2} \epsilon_{jj'} \nu_{jj'}$, where $j$ and $j'$ index the states and $j \neq j'$, $\epsilon_{jj'}$ permutes $j$ and $j'$, $\mu = k_x, \xi_z$. The term

$$\nu_{jj'} = \oint \nabla_{\vec{\mu}} \text{Arg}[\omega_j(\vec{\mu}) - \omega_{j'}(\vec{\mu})] \cdot d\vec{\mu} \tag{8}$$

is called the eigenvalue vorticity[41,45]. By encircling an EP on the WER as shown in Fig. 2e, we find $Y = 1$.

In addition, driving the eigenstates (Bloch wavefunctions) around the same loop (Fig. 2e) produces a non-Hermitian Berry phase given by

$$\theta = \oint i \langle \psi_m(\vec{\mu}) | \partial_\mu \psi_m(\vec{\mu}) \rangle \cdot d\mu. \tag{9}$$

Because the evolution of eigenstates is bounded to the non-Hermitian manifold that has two eigenvalue sheets connected at the branch cut, two complete cycles along the encircling path are required for both eigenstates to recover. The total Berry phase after two cycles is $\theta = \pi$[27]. Hence the eigenstate winding number is $W = 1/2$.

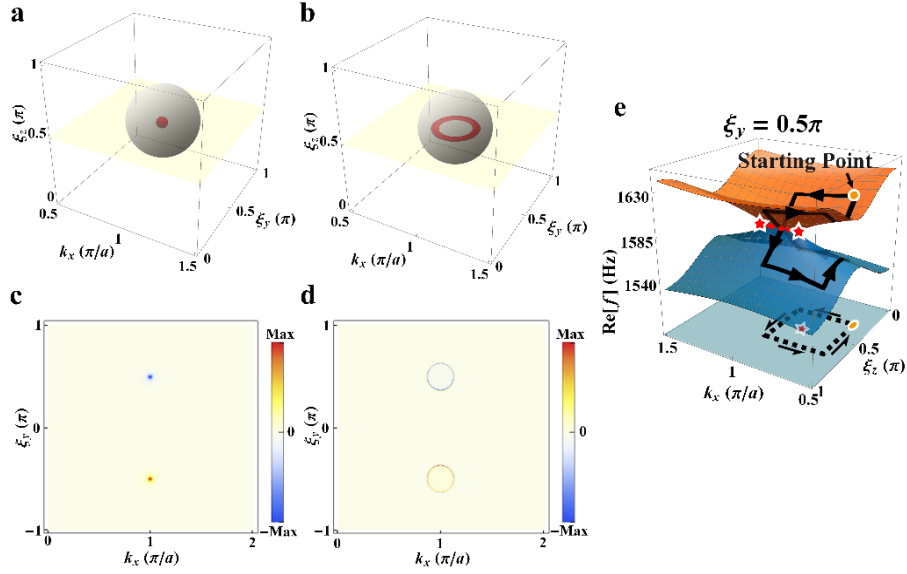

**Fig. 2 Schematics of calculations of Hermitian and non-Hermitian topological charges.** The gray surfaces enclose **a** the WP and **b** the WER for the calculation of topological charge. And the yellow plane represents the $k_x\xi_y$ plane at $\xi_z = 0.5\pi$. **c, d** The distribution of the Berry curvature of Hermiticity and non-Hermiticity in the $k_x\xi_y$ plane at $\xi_z = 0.5\pi$, respectively. The color scale indicates the magnitude of the Berry curvature. **e** Schematics of eigenfrequency trajectories for looping around the EP in the $k_x\xi_z$ plane at $\xi_y = 0.5\pi$, which are used to calculate a quantized Berry phase characterizing the EP. The bottom is a projection of the loop on the $k_x\xi_z$ plane.

*Experimental results.* The non-Hermitian model is pictorially shown in Figs. 3a, b. Experimentally, we construct a 1D finite PC with $N = 24$ metal cavities and a circuit part shown in Fig. 3c. Onsite orbitals are mimicked by the first-order cavity resonance. The pressure profile of the onsite mode is shown at the upper-right panel in Fig. 3c. Because the eigenfrequency of cavities can be adjusted by the height of cavities, the synthetic dimension $\xi_z$ associated with the onsite modulation described by Eq. (2) can be straightforwardly implemented (see Supplementary Table 1). The cavities are connected by small tubes to implement the tight-binding hopping $\kappa(k_x, \xi_y)$. The Bloch wavenumber $k_x$ is naturally realized by the periodicity of the PC. At the same time, by further modulating the tubes' cross-sectional areas according to Eq. (3), the synthetic dimension $\xi_y$ is realized (see Supplementary Table 2). The non-Hermiticity in Eq. (1) is introduced as controlled additional loss and gain. To achieve this, an in-house-designed active unit comprised of a loudspeaker, a microphone and a feedback circuit is installed at the top of each cavity. The amplitude and phase of the emission are precisely controlled according to the signal measured by the microphone. Gain and loss are introduced when the emission of the loudspeaker is in-phase and anti-phase with respect to the sound at the top of the cavity, respectively (see Supplementary Note 2 for more details).

We begin with a simple two-cavity setup to obtain the system parameters of the acoustic system. By using the Green's function approach and a least-square fitting method[42,46,47], the system parameters are extracted to be $\omega_0 = 9953.8$ rad/s, $\kappa_0 = 118.7$ rad/s, $\gamma_0 = 30.8$ rad/s, $\gamma_1 = 67.8$ rad/s and $\gamma_2 = 33.9$ rad/s. More details about the two-cavity experiments and fitting procedures are shown in Supplementary Note 3.

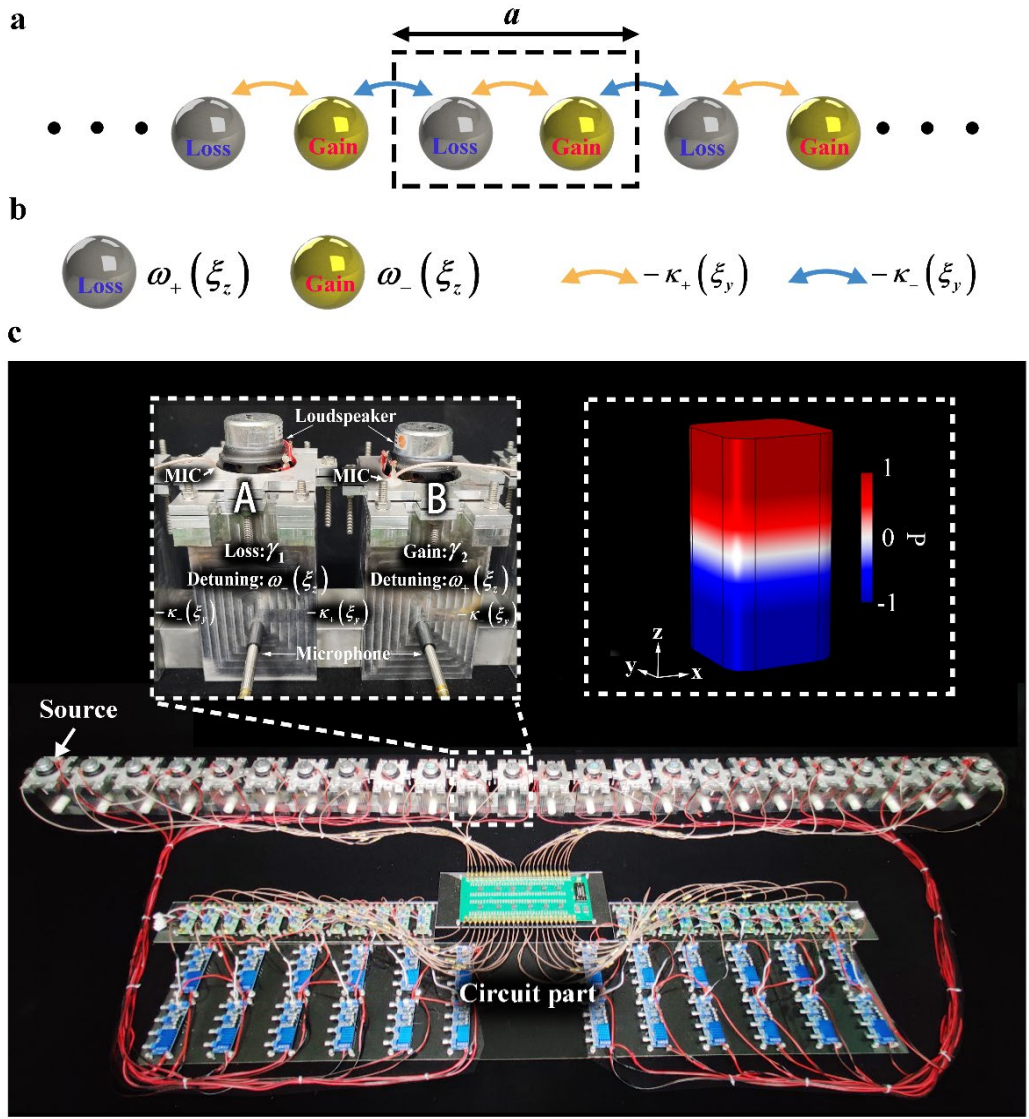

**Fig. 3. Theoretical model and corresponding experimental system. a** The 1D non-Hermitian model. The dotted box indicates the unit cell. **b** The loss and gain are introduced to onsite terms, which are modulated by functions $\omega_\pm(\xi_z)$, and the coupling strength of intra-cell and inter-cell are modulated by functions $-\kappa_\pm(\xi_y)$. **c** The corresponding experimental system is composed of the cavities part and the circuit part. The circuit part connecting the loudspeakers and microphones at the top of cavities is to actively control the energy in the cavities for introducing additional loss and gain. The left inset shows the zoomed-in view of two cavities corresponding the experimental unit cell ($a$ = 210 mm). The pressure profile $P$ of the cavity's mode is shown at the right.

Experimentally, we place an acoustic source at the left-most cavity of the PC and measure the pressure response spectra in each cavity with a microphone. Fourier transform is then applied to convert the spatial coordinate $x$ to reciprocal coordinate $k_x$. First, we verify the existence of the Weyl point. The two synthetic coordinates are set at $\xi_y = \xi_z = 0.5\pi$, as indicated by the grey cut plane at left of Fig. 4a. The measured dispersion relation (real part of the frequency) is shown at the left of Fig. 4b, wherein a continuous band is seen. This band is linear in the vicinity of $k_x = \frac{\pi}{a}$, which agrees well with the theoretical prediction (green line) using the designed Hamiltonian and the parameters extracted from the two-cavity setup. Next, the non-Hermiticity is introduced by turning on all the active units, which are set to staggeringly generate the gain and loss. In the measured dispersion shown at the right of Fig. 4b, a flat plateau is developed near $k_x = \frac{\pi}{a}$. This plateau corresponds to the PT-broken phase, at which the real parts of the eigenfrequencies are degenerate. The two endpoints of this plateau are two EPs on the WER. We further obtain the dispersion at 11 different $\xi_y$ with $\xi_z = 0.5\pi$, as shown in Fig. 4c. The dispersion curves are gapped for $\xi_y < 0.37$ and $\xi_y > 0.63$. The plateau can be clearly identified for $0.44 \leq \xi_y \leq 0.56$. The plateau's width increases until $\xi_y = 0.5$ and then decreases and eventually vanishes at $\xi_y = 0.63$. The endpoints of the plateau clearly delineate a closed loop on the $k_x\xi_y$ plane at $\xi_z = 0.5\pi$, which validate the existence of the WER. We repeat the measurement with $\xi_y = 0.5\pi$ and then tune $\xi_z$ to 7 different values. In the results plotted in Fig. 4d, we can identify a gap that closes only at $\xi_z = 0.5\pi$ in the formation of a plateau. From this observation, we conclude that the WER indeed lies on the $k_x\xi_y$ plane at $\xi_z = 0.5\pi$.

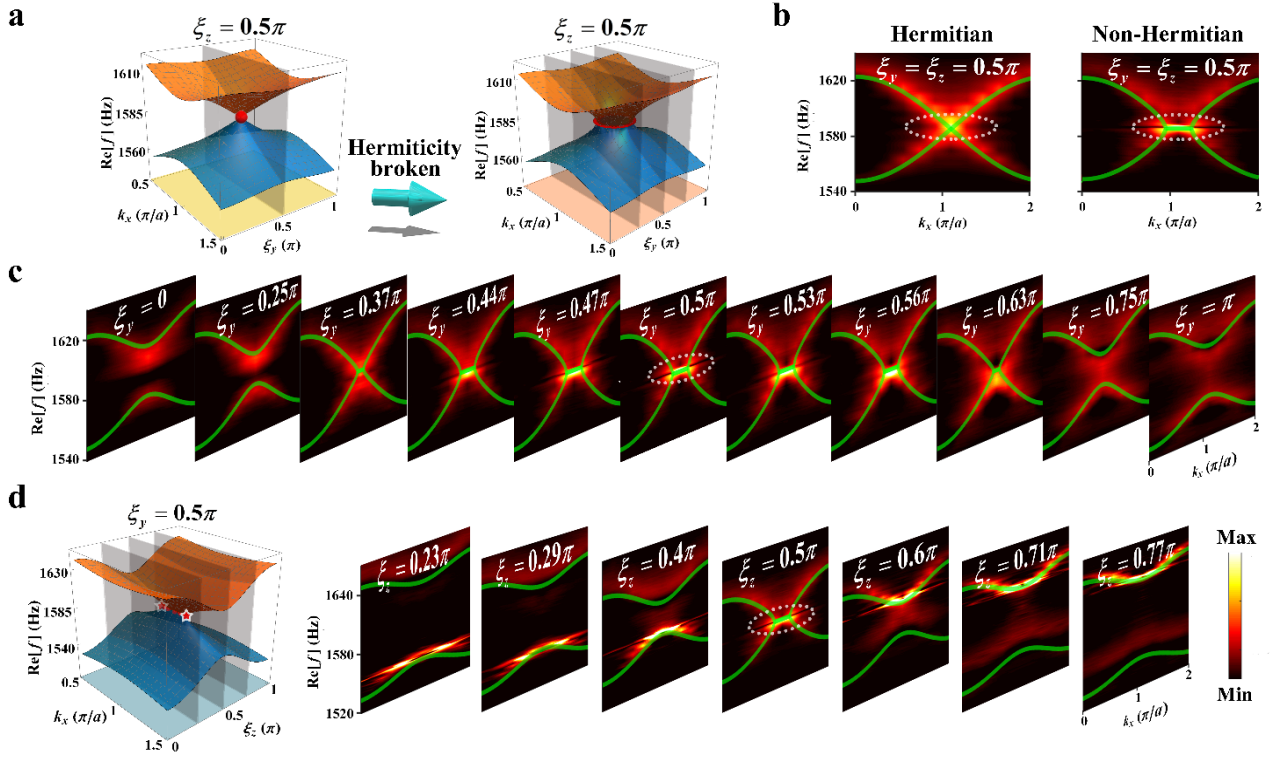

**Fig. 4 Experimental results. a** Schematic of using different planes to cut two band structures. **b** Measured dispersion of the 1D PC. After we add additional loss and gain in the cavities, the point evolves into a nodal line at eigenfrequency $\omega_0$ at $\xi_y = \xi_z = 0.5\pi$. The green lines correspond to the numerically calculated band structures of the designed Hamiltonian. **c** For different values of $\xi_y$ at $\xi_z = 0.5\pi$, the images show the evolution of band structures. **d** is similar to **c**, but it is for different values of $\xi_z$ at $\xi_y = 0.5\pi$. We can deduce that the WP becomes a WER and exists in the $k_x\xi_y$ plane at $\xi_z = 0.5\pi$ in the synthetic momentum space.

The presence of the WPs indicates the existence of Fermi arcs which are 2D surface states pinned by a pair of WPs with opposite topological charges in a finite-sized system. In our system, the Fermi arcs are observed as 0D topological boundary modes localized at the two ends of the real-space PC. Fermi arcs are generally dispersive curves in the reciprocal dimensions. But here, due to the inversion symmetry[34], they become zero-energy flat bands that are straight segments composed of edge states at between $-0.5\pi < \xi_y < 0.5\pi$ at $\xi_z = 0.5\pi$, as shown in Fig. 5a. In the presence of non-Hermiticity, the WPs become WERs. Because the inversion symmetry is preserved, the Fermi arc remains a straight segment that connects the WERs shown in Fig. 5b. We have measured the Fermi arcs composed of edge states experimentally for both the Hermitian and non-Hermitian cases at a series of $\xi_y$ with $\xi_z = 0.5\pi$. The results are plotted in Figs. 5c, d as blue markers, which confirm that the Fermi arcs are indeed zero-energy flat bands.

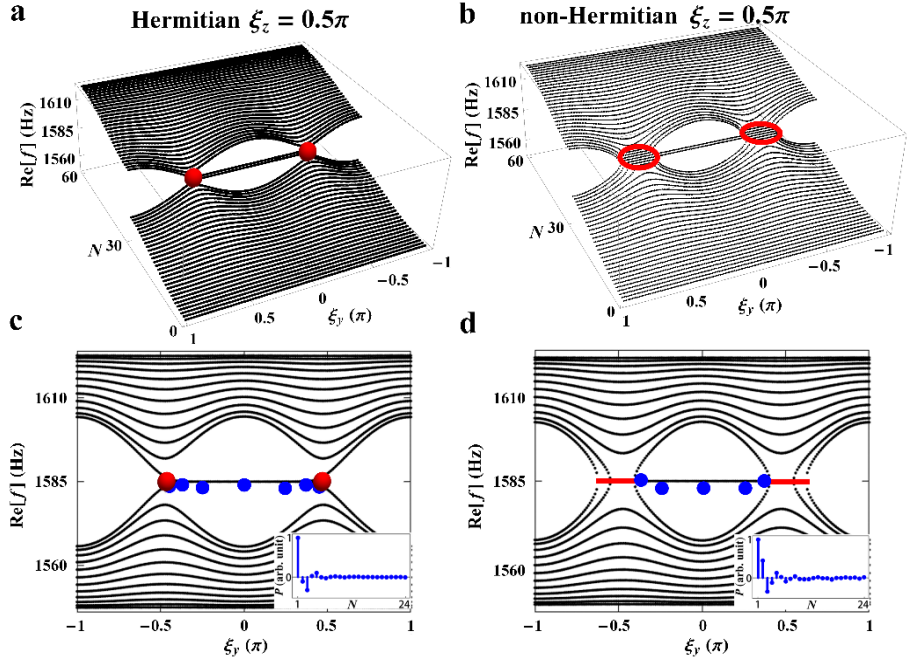

**Fig. 5 Fermi arc of WPs and WERs.** The real spectra of a finite lattice ($N = 60$) with an open boundary condition and $\xi_z = 0.5\pi$ of the Hermitian case **a** and non-Hermitian case **b**. The middle chain is composed of edge states. **c, d** The projected band structures ($N = 24$), in which the blue dots are edge states of experimental results that indicate the trajectories of the Fermi arcs with Hermitian condition and non-Hermitian condition, respectively. The insets show the measured of the pressure field of Fermi arcs at $\xi_y = 0$.

***Conclusion***. We have experimentally realized a WP and a WER in a 1D non-Hermitian PC with two synthetic dimensions. Our results show that EP structures appearing in the Bloch bands of non-Hermitian systems can be investigated in hybrid reciprocal-synthetic systems, which are experimentally more convenient and versatile compared to systems with pure spatial periodicity. We have also developed an acoustic system with the successful implementation of actively tunable loss and gain, which can serve as a platform for studying more sophisticated non-Hermitian phenomena, such as non-Hermitian skin effects[24,48], EP chains[42], EP links[49].

*Acknowledgments.* This work was supported by National Key R&D Program of China (Grant No. 2017YFA0303700), the National Natural Science Foundation of China (Grant Nos 11634006, 11374157, 81127901, 11802256, 11922416), Hong Kong Research Grant Council (12302420, 12300419, 22302718, C6013-18G), a project funded by the Priority Academic Program Development of Jiangsu Higher Education Institutions, the Innovation Special Zone of National Defense Science and Technology and High-Performance Computing Center of Collaborative Innovation Center of Advanced Microstructures.

*Author Contributions.* Bin Liang and Guancong Ma conceived the research; Guancong Ma, Ze-Guo Chen and Zheng-wei Li performed theoretical analysis; Zheng-wei Li, and Jing-jing Liu carried out numerical calculations and experiments; Zheng-wei Li, Jing-jing Liu and Ze-Guo Chen made equal contributions to this research; all authors analyzed and discussed the results and contributed to the manuscript.